\documentstyle[11pt,aaspp4]{article}      
\def\ecs{erg cm$^{-2}$ s$^{-1}$~}
\def\hes{h$_{50}^{-2}$ erg s$^{-1}$~}
\def\qo{q$_{\circ}$}
\def\chis{$\chi^2$}

\lefthead{Jones et al.}
\righthead{X-ray evolution of low luminosity clusters}

\begin{document}                        

\title{The WARPS survey - II: The Log(N)-Log(S) relation
and the X-ray evolution of low luminosity clusters of
Galaxies}

\author{L.R. Jones\altaffilmark{1,2,7,9}, C. Scharf\altaffilmark{2,8}, 
H. Ebeling\altaffilmark{3,9}, E. Perlman\altaffilmark{2,4,7,8},
G. Wegner\altaffilmark{5}, M. Malkan\altaffilmark{6} and 
D. Horner\altaffilmark{2,8} }

\altaffiltext{1}{School of Physics and Space Research, University of
Birmingham,
Birmingham B15 2TT, UK. Email: lrj@star.sr.bham.ac.uk}
\altaffiltext{2}{Laboratory for High Energy Astrophysics, Code 660,
NASA/GSFC, Greenbelt, MD 20771, USA.}
\altaffiltext{3}{Institute for Astronomy, 2680 Woodlawn Dr, Honolulu, HI
96822, USA}
\altaffiltext{4}{Space Telescope Science Institute, Baltimore, MD 21218,
USA.}
\altaffiltext{5}{Dept. of Physics \& Astronomy, Dartmouth College, 6127
Wilder Lab., Hanover, NH 03755, USA.}
\altaffiltext{6}{Dept. of Astronomy, UCLA, Los Angeles, CA 90024, USA}
\altaffiltext{7}{Visiting Astronomer, Kitt Peak National Observatory,
National Optical Astronomy Observatories, which is operated
      by the Association of Universities for Research in Astronomy, Inc.
(AURA) under cooperative agreement with the
      National Science Foundation.}
\altaffiltext{8}{Visiting Astronomer, Cerro Tololo Interamerican
Observatory, National Optical Astronomy Observatories, which is operated
      by the Association of Universities for Research in Astronomy, Inc.
(AURA) under cooperative agreement with the
      National Science Foundation.}
\altaffiltext{9}{Visiting Astronomer, Canada-France-Hawaii Telescope,
operated by the
     National Research Council of Canada, the Centre National de la
     Recherche Scientifique de France and the University of Hawaii.}

\begin{abstract} 
The strong negative evolution observed in previous X-ray selected surveys
of clusters of galaxies is evidence in favour of hierarchical models 
of the growth of structure in the Universe. A large recent survey
has, however, contradicted the low redshift results, finding no evidence for 
evolution at z$<$0.3 (Ebeling et al.\markcite{e97a} 1997a). Here we
present the first results from an X-ray selected, flux and 
surface brightness limited 
deep survey for high redshift clusters 
and groups of galaxies based on ROSAT PSPC pointed data.
The log(N)-log(S)
relation of all clusters in this survey is consistent with that from
most previous surveys but occupies a flux range not previously covered 
($>$6x10$^{-14}$ \ecs total flux in the 0.5-2 keV band).
At high redshifts (z$>$0.3) the cluster luminosities are in the range
4x10$^{43}$ \hes to 2x10$^{44}$ \hes, the luminosities of poor
clusters. The number of high redshift, low luminosity
clusters is consistent with no evolution of the X-ray luminosity 
function between redshifts of z$\approx$0.4 and z=0, and places
a limit of a factor of $<$1.7 (at 90\% confidence) 
on the amplitude of any pure negative density evolution
of clusters of these luminosities, in contrast with the factor of
$\approx$3
(corresponding to number density evolution $\propto$(1+z)$^{-2.5}$)
found
in the EMSS survey  at similar redshifts but higher
luminosities. Taken together, these 
results support hierarchical models
in which there is mild negative evolution
of the most luminous clusters at high redshift but little or no evolution
of the less luminous but more common, optically poor clusters. 
Models involving preheating of the X-ray gas at an early epoch fit
the observations, at least for $\Omega_0$=1.
\end{abstract}


\section{Introduction}

Measuring the evolution of clusters of galaxies is a powerful test of
hierarchical models of the gravitational growth of structure in the
Universe. The most massive clusters are rare and in many models (e.g. the
Cold Dark Matter model) the majority are predicted to have formed in the
relatively recent past via the merger of less massive clusters. The rate
of evolution of the properties of the cluster population, such as the
X-ray luminosity and temperature functions, over a wide range of cluster
masses, can help discriminate between different model parameters and
between different thermal histories of the dominant X-ray gas.

X-ray surveys of clusters have the advantage in principle of being
relatively unbiased, since they are unaffected by projection effects.
Indeed, the detection of diffuse X-ray emission represents direct
evidence of a deep gravitational potential within which the hot
intra-cluster gas is trapped. Furthermore, the X-ray properties of the
hot gas can be directly related to the gravitating mass, inferred to be
dominated by a dark component. However, the observational evidence for
X-ray evolution has not always been consistent in detail, even from
purely X-ray selected and X-ray flux limited samples. One of the first
determinations of the local X-ray luminosity function (XLF) using an
X-ray selected sample was by Piccinotti et al.\markcite{p82} (1982),
using non-imaging detectors. The first claims of a measurement of
evolution in the cluster XLF were by Edge et al.\markcite{ed90} (1990)
and Gioia et al.\markcite{g90} (1990). Edge et al. compiled a list of 46
clusters and concluded that strong negative evolution was observed, with
the number density of clusters of high luminosity (L$_X>$10$^{45}$
h$_{50}^{-2}$ erg s$^{-1}$) increasing by a factor of $\sim$10 over the
redshift range z=0.18 to z=0. From the 67 clusters at z$>$0.14 imaged in
the {\it Einstein} Extended Medium Sensitivity Survey (EMSS), Gioia et
al.\markcite{g90} (1990) and Henry et al.\markcite{h92} (1992) found
evidence (at 3$\sigma$ significance) for a lower rate of evolution: for
example the number density of high luminosity clusters
(L$_X\approx$6x10$^{44}$ h$_{50}^{-2}$ erg s$^{-1}$) was found to
increase by a factor of $\approx$5 between median redshifts of z=0.33 and
z=0.17. Castander et al.\markcite{c95} (1995) found that the number
density of lower luminosity clusters (L$_X\gtrsim$1x10$^{43}$
h$_{50}^{-2}$ erg s$^{-1}$) also showed evolution, with a factor of
$\approx$2 increase from the redshift range 0.2$<$z$<$0.55 to z=0
(although we find a different result in this paper; see section 5.5).

Recent results have altered this picture of strong evolution
dramatically. The first of several large cluster samples being derived
from the ROSAT All-Sky Survey (the ``brightest cluster sample'' or BCS)
contains $\approx$200 clusters and shows {\it no} evidence for evolution
of the XLF at low redshifts, with a change in normalisation of the XLF of
a factor of $\lesssim$1.6 (at 68\% confidence; Ebeling et
al.\markcite{e97a} 1997a fig. 5) for luminosities of L$_X>$4.5x10$^{44}$
h$_{50}^{-2}$ erg s$^{-1}$ between median redshifts of z=0.21 and z=0
(Ebeling et al.\markcite{e97b} 1997b). Ebeling et al.\markcite{e95}
(1995, 1997a) show that this inconsistency with the results of Edge et
al.\markcite{ed90} (1990) is due to the small sample size and unfortunate
sampling in redshift space (together with a volume miscalculation) in the
Edge et al.\markcite{ed90} sample, and that the rate of evolution at
z$<$0.3 measured from a much larger sample is considerably smaller than
previously thought.

At higher redshifts, the results of the very recent survey of Collins
\markcite{c97}
et al. (1997) contradict those of Castander et al.\markcite{c95} (1995).
Collins et al.\markcite{c97} find that the number of low luminosity
clusters at z$>$0.3 shows no evolution, even though the Collins et
al.\markcite{c97} survey was not complete for the most extended X-ray
sources. The EMSS sample of Henry et al.\markcite{h92} (1992) has been
reanalysed by Nichol et al.\markcite{n97} (1997) who replaced {\it
Einstein} IPC fluxes with ROSAT PSPC fluxes for 21 clusters and discarded
7 objects as unlikely to be clusters (although this aspect relied heavily
on whether the objects were resolved in the ROSAT PSPC). Nichol et
al.\markcite{n97} still found evidence for evolution of the XLF but at a
lower rate than that measured by Henry et al.\markcite{h92} At even
higher redshifts, Luppino \& Gioia\markcite{l95} (1995) found no evidence
in the EMSS for further evolution between 0.6$<$z$<$0.8 and
z$\approx$0.33 for clusters of similar luminosity
(L$_X\approx$6x10$^{44}$ h$_{50}^{-2}$ erg s$^{-1}$), although their
small sample size meant that a factor of $\approx$2 in number density
evolution was allowed between z$\approx$0.33 and z$\approx$0.7. It is
worth noting that, despite apparently contrary claims about the presence
of strong evolution, the EMSS XLF agrees well with that Ebeling et
al.\markcite{e97a} 1997a where the two samples overlap in redshift. The
evolution seen in the EMSS is limited to z$>$0.3 and thus not in conflict
with the low redshift results.

Thus, the recent X-ray results suggest that the evolution of the
luminosity function of clusters is less rapid than previously thought,
but that there is still evidence for evolution of X-ray luminous systems
at high redshifts (z$\gtrsim$0.3). In contrast, optical surveys for
distant (z$>$0.3) clusters have found the number density of rich clusters
at high redshifts to be approximately the same as measured locally (Gunn
et al.\markcite{g86} 1986, Couch et al.\markcite{c91} 1991, Postman et
al.\markcite{p96} 1996). This difference may be due to the highly
non-linear dependance of the X-ray 
luminosity on mass, so that a small change in mass (and richness) results in 
a large change in luminosity. 

Current X-ray selected and X-ray flux limited samples contain few
clusters at high redshifts, and even fewer high redshift, low X-ray
luminosity clusters. Here we describe the first results from the WARPS
(Wide Angle ROSAT Pointed Survey) cluster/group survey. This X-ray
selected, X-ray flux limited survey was designed primarily to measure the
high redshift (z$>$0.3) cluster XLF at lower luminosities than the EMSS
(L$_X\gtrsim$3x10$^{43}$ h$_{50}^{-2}$ erg s$^{-1}$), but it also
contains groups of galaxies, which have lower luminosities than clusters
and are therefore detectable at lower redshifts of z$\approx$0.1, and nearby
individual galaxies which have been resolved. In this paper we
concentrate on the evolution of clusters of galaxies of 
L$_X>$3x10$^{43}$ h$_{50}^{-2}$ erg s$^{-1}$. We assume that
groups and clusters of galaxies form a continuous population, referring
to the population simply as `clusters', and do not further distinguish
groups of galaxies from clusters. 
Future papers will investigate the detailed properties of
all these systems. The survey design places particular emphasis on a high
level of completeness in both X-ray source detection and cluster
identification.  Our application of the X-ray source detection technique
(VTP or Voronoi Tessellation and Percolation), the source classification
and the survey calibration are described in Scharf et al.\markcite{s97} (1997)
(hereafter Paper I). Based on a larger sample for which optical
identifications are currently being obtained, a future paper will
describe the WARPS cluster XLF. Here we present the X-ray log(N)-log(S)
relation (i.e. the number of clusters as a function of flux) for the
current, statistically complete sample of confirmed clusters, both at all
redshifts and at high redshifts alone, and use it to constrain the
evolution of the cluster XLF.

In Section 2 we describe the sample selection. The optical observations
are described in Section 3, and the 
the log(N)-log(S) relations are presented in Section 4. In Section 5
a comparison is made with the predictions of various models of
the growth of structure in the Universe. An appendix gives details 
of the X-ray K-corrections used.
Unless otherwise stated, we use q$_{\circ}$=0.5 and H$_{\circ}$=50 h$_{50}$ km
s$^{-1}$ Mpc$^{-1}$.

\section{The Sample}

Our sample is based on ROSAT Position Sensitive Proportional Counter
(PSPC) X-ray data from 86 pointings with exposures $>$8 ks (up to 48 ks)
and galactic latitude $|$b$|>20^{\circ}$. We set a limit of
3.5x10$^{-14}$ erg cm$^{-2}$ s$^{-1}$ in detected flux within the energy
range of 0.5-2 keV. The observed redshift range of clusters is from z=0.1
to z=0.67 with a mean redshift $\approx$0.25; X-ray luminosities range
from 1x10$^{42}$ h$_{50}^{-2}$ erg s$^{-1}$ to 2x10$^{44}$ h$_{50}^{-2}$
erg s$^{-1}$ (0.5-2 keV).

We minimize the Galactic contribution to the X-ray background by
selecting a lower bound to the bandpass of 0.5 keV. Importantly, this
also minimizes the size of the instrumental point spread function (PSF),
while maintaining a high signal from gas at the temperatures found in
clusters of galaxies. We use the part of each PSPC X-ray image within
radii of 3 arcmin to 15 arcmin, avoiding the target of the pointing at
low radii and the shadow of the window support structure which moved with
the (deliberate) spacecraft wobble at large radii. The instrumental PSF
also degrades rapidly at off-axis angles $>$15 arcmin. The original
targets of the PSPC observations were nearly all Active Galactic Nuclei
(AGN), stars or nearby galaxies. Five of the 86 observation targets were
clusters or groups of galaxies, which could introduce a small bias, since
clusters cluster amongst themselves. However, in none of these fields was
a serendipitous cluster found at a redshift near that of the original
target (within $\Delta$z=0.1), and below we show that the conclusions of
the paper are not affected if these five fields are ignored. The
non-cluster extragalactic targets could in principal introduce a small
bias, if, for example, some fraction were AGN in a supercluster. An
initial check shows that the fraction of fields with extragalactic
targets containing serendipitous clusters above our flux limit
(40$\pm$9\%) is not significantly different from the fraction with
galactic targets (33$\pm$9\%). We note that here there are 5 fewer fields
in total (86 rather than 91) than described in Paper I. This is because
very bright stars or large nearby galaxies were found to mask a large
fraction of these fields. The total survey sky area is 16.2 deg$^{2}$.

Here we only summarize the source detection and classification procedure,
since a full description is given in Paper I. Each PSPC field is
corrected for non-uniform exposure and vignetting using energy dependent
exposure maps. The source detection algorithm is Voronoi Tessellation and
Percolation, described by Ebeling and Wiedenmann\markcite{e93} (1993)
and in Paper I. The algorithm is very general, not preferentially
detecting sources of any particular size or shape. An isophotal threshold
in X-ray surface brightness a small factor (typically 1.4) above the
background level is computed for each field. The individual sources are
formed by grouping together neighbouring photons that lie above the
surface brightness threshold. To help separate close sources which can be
combined incorrectly into one source, the algorithm is rerun using 3-5
increasing threshold levels, and the final source catalogue compiled
using the results from all thresholds.  The $10^{\rm th}$ and $90^{\rm
th}$ percentiles of the local thresholds for our final source list are
1.5 (for very extended, faint sources) and 3.0 (for deblended point sources),
respectively.

Knowing the surface
brightness threshold used for each source, the counts above the
threshold, and the sky area in which they were detected, the total count
rate extrapolated to infinite radius is calculated for each source 
assuming that the source profile is given by (a)
the position-dependent PSF only, and (b) the PSF convolved with the best fit
King profile. We assume that $\beta$=$\frac{2}{3}$, the average value
found
by Jones \& Forman\markcite{j84} (1984), and measure the angular core 
radius (Ebeling et al.\markcite{e96} 1996, 1997b). A source is classified as
extended if the ratio of the total fluxes calculated using the two
assumptions exceeded a critical value 
determined from simulations (see Paper I). 

A conversion from count rate to (absorbed) flux in the 0.5-2 keV band was
performed using a constant factor of 1.15x10$^{-11}$ \ecs (ct
s$^{-1}$)$^{-1}$. The maximum Galactic equivalent column density of
neutral hydrogen ($N_H$) in the direction of our fields is 1.4x10$^{21}$
cm$^{-2}$, and 90\% of the fields have $N_H$ in the range from
9x10$^{19}$ cm$^{-2}$ to 7x10$^{20}$ cm$^{-2}$. For this range of column
density, and abundances of 0.25 times the cosmic abundance, even with
Raymond \& Smith\markcite{r76} (1977) spectrum temperatures of 1.4 keV to
14 keV the constant flux conversion factor is accurate to within 6\%, and
thus no correction for absorption variations has been made. The constant
correction to unabsorbed fluxes (i.e. removing the effect of Galactic
absorption) was made using a factor of 1.1, corresponding to the median
$N_H$ of 3.5x10$^{20}$ cm$^{-2}$. This factor is almost independent of
temperature and varies by $\pm$10\% within the above temperature and
$N_H$ ranges.

The correction from detected flux to total flux (i.e. extrapolated to
infinite radius, but remaining in the 0.5-2 keV band) for extended
sources which have been confirmed as clusters is typically a factor of
1.4 (but is computed for each source separately). A plot of total flux
versus detected flux is shown in Figure 1 for all candidate clusters. A
few point-like sources, for which the flux correction is small, are
clearly visible close to the dashed line defining zero correction. These
are cluster candidates which have been identified via our optical imaging
program of point-like sources. The survey is complete to a flux limit in
total flux of 6x10$^{-14}$ \ecs (0.5-2 keV), higher than the flux limit
of 3.5x10$^{-14}$ \ecs (0.5-2 keV) in detected flux (shown by the dotted
lines). The measured core radii of resolved sources are typically in the
range 0.3 arcmin to 0.6 arcmin. Simulations, shown in Figures 5 \& 6 of
Paper I show that in this range of core radii the total flux is recovered
to within 10\% accuracy for all signal to noise ratios and off-axis
angles used, at least for the well-behaved King profiles used in the
simulations. For a typical high redshift cluster in the survey at z=0.5
with a luminosity of L$_X\approx$1x10$^{44}$ h$_{50}^{-2}$ erg s$^{-1}$,
a core radius of 0.40 arcmin corresponds to r$_c$=170 h$_{50}^{-1}$ kpc
(\qo=0.5). Since this core radius is in reasonable agreement with those
measured for nearby clusters, we are confident that the total count rates
for most of our clusters, or at least those which are well described by a
King profile, are accurate to within 10-20 per cent.

The sky area in which a source of a given total flux and intrinsic 
core radius could have been detected (including point sources)
has been calculated via a combination of  simulations and 
an analytical approach, as 
described in Paper I. The different
exposure and background level of each PSPC field, and the position 
dependent PSF are all taken into account. The fraction of the total
survey area available as a function of total flux and intrinsic core
radius is given in Figure 8 of Paper I.  In
practice few sources of large angular size (core radius $>$0.7 arcmin) 
have been detected, although the survey was sensitive to them,
and most of the extended sources, with core radii in the range 0.3 to 0.6
arcmin, could have been detected within $>$90\% of the 
total survey area. In Section 5.4 below, we estimate how many large,
very low surface brightness sources we expect in our survey, and 
find that the survey was sensitive to nearly all the sources predicted
above the flux limit - i.e. the survey was nearly completely flux limited
rather than surface brightness limited.

\section{Optical observations}

Here we describe the method used to categorize the optical counterparts
and the action taken in the optical follow-up program. Because most high
latitude X-ray sources at the fluxes considered here are AGN (e.g. Shanks
et al.\markcite{s91} 1991), we select cluster and group candidates for
spectroscopy based on the X-ray extent, sky survey plate measurements and
CCD imaging.

Although clusters of core radii of 7 arcsec can be resolved on axis if
the signal-to-noise ratio of the PSPC X-ray data is high, a more
realistic limit, including off axis data, is $\approx$20 arcsec (see
Figure 7 of Paper I), which corresponds to 140 h$_{50}^{-1}$ kpc at
z=0.5. This resolution is adequate to resolve most clusters and groups
with average core radii for their luminosity at the redshifts we expect
to detect them (e.g. the mean core radius found by Jones \& Forman
\markcite{j84} 1984 for low redshift clusters was 250 h$_{50}^{-1}$ kpc).
However, clusters have a wide range of morphologies (even within a small
range of X-ray luminosity) and cooling flow clusters, unusually compact
systems, or those which contain both extended emission and point sources
could be classified erroneously as point-like (see Evrard \&
Henry\markcite{ev91} 1991).  Edge et al. \markcite{e92} (1992) measured
substantial cooling flows ($>$100 M$_{\sun}$ yr$^{-1}$) in 23\% (5 of 22)
clusters with luminosity $<$3x10$^{44}$ erg s$^{-1}$, indicating that
cooling flows may be relatively common even in low luminosity systems.
Cooling flows produce a peaked X-ray surface brightness profile. For
instance, Nichol et al.\markcite{n97} (1997) report that an HRI image of
the luminous EMSS cooling flow cluster MS2137.3-2353 at a redshift of
z=0.313 gives a core radius of 17$\pm$8 arcsec, corresponding to 95
h$_{50}^{-1}$ kpc. MS1512.4+3647, at a redshift of z=0.373, has an even
smaller core radius of 7$\pm$1.5 arcsec (Hamana et al. \markcite{ha97}
1997). Sources with core radii this small may not be resolved in the PSPC
(depending on off-axis angle), and therefore to maximise the completeness
we include in the spectroscopic follow-up both extended sources
regardless of their optical counterparts {\it and} point-like X-ray
sources which have an excess of galaxies on R band CCD images. We do not
include point-like X-ray sources which have only stellar optical
counterparts (AGN and stars).

First, APM machine measurements of Palomar E and O and UKST R and B$_j$
plates are obtained at the positions of all X-ray sources of flux
$>$3.5x10$^{-14}$ erg cm$^{-2}$ s$^{-1}$ (0.5-2 keV) at off-axis angles
$<$15 arcmin. This gives typically 3-4 sources per field plus the target
of the observation. The systematic PSPC pointing error, which is $\sim$15
arcsec in size and varies in direction between observations (Briel et
al.\markcite{b95} 1995), was removed by inspection of the optical maps at
the positions of point X-ray sources, including the target of the
observation where available. In nearly all fields the pointing error
could be immediately determined to within $\approx$5 arcsec, since most
of these X-ray sources have a single optical counterpart with a similar
offset from the X-ray position as the target. The mean offset of these
sources is taken as the X-ray pointing error. Possible optical
counterparts with magnitudes near the plate limits are ignored in this
procedure. The remaining random position errors for point X-ray sources
are of mean (and 95\%) size 4.7$\pm$0.6 (9.7) arcsec. These error circle
sizes were confirmed during the spectroscopic follow-up. A sample of 21
spectroscopically confirmed AGN has a mean (and 95\%) position error of
4.8$\pm$0.6 (9.3) arcsec. An error circle radius of 10-15 arcsec was
adopted, depending on the signal-to-noise ratio of the detection.

For all sources, the X-ray contours are overlaid on digitized versions of
the optical plate material in search of obvious optical counterparts.
Depending on whether the X-ray source is extended or not, we then proceed
as follows:

If, for extended X-ray sources, there is an excess of bright
(R$<$19 mag) galaxies within the X-ray contours,
optical spectra are obtained of between 2 and 6  galaxies. If at least
2 galaxies (in the case of 2 or 3 spectra) or at least
3 galaxies (in the case of 4 or more spectra) have
very similar redshifts, the source is identified as a cluster. 
CCD R band images of most of these clusters have been obtained. 
If the redshifts are not similar, or there is no excess of galaxies on
the plate,
imaging to R=23 mag (or to R=24.5 mag or in the I band in some cases) 
is obtained,
objects selected for spectroscopy, and the process repeated. In general,
the objects selected for spectroscopy in cluster candidates
are not only  the brightest galaxies
but also those objects (including stellar objects)
near peaks in the X-ray surface brightness. This process is important
in determining the fraction of X-ray emission {\it not} from the intra-cluster 
medium, and also in cases where no excess
of galaxies are found to R$\sim$24 mag at the position of an extended 
X-ray source. In these latter cases we have so far found that 
in each case the X-ray source is not truly extended, but a blend
of several very close
point-like sources, and the counterparts include AGN and stars.

For point X-ray sources, the APM magnitudes and source extent measurement
(the `stellarness' parameter; Irwin, Maddox \& McMahon\markcite{i94}
1994) of the optical counterpart(s) are used to define the next action.
If the error circle is blank, imaging to R=22 mag or fainter is obtained.
If the error circle contains only faint APM objects, within 1 magnitude
of the plate limit, then the APM source extent measurement is assumed to
be unreliable and again R band imaging is obtained. In addition, if the
X-ray extent parameter is in the range 1.1-1.2, just below the critical
value of 1.2 above which a source is considered to be extended, CCD
imaging is also obtained, regardless of the content of the error circle.
If the CCD image contained an excess of galaxies in or close to the error
circle, spectra were obtained of the galaxies as well as objects within
the error circle. More usually, there was a single counterpart in the
error circle. The FWHM of the counterpart was compared to stars on the
same CCD image, and if it was at least 3$\sigma$ greater than the mean
stellar FWHM, the source was designated as a galaxy.

However, most ($\approx$70\%) of the point X-ray sources contained a
single counterpart within the error circle on the Palomar E or UKST R
plates, and $\sim$70\% of these were bright enough to have an accurate
APM `stellarness' measurement (defined as R$<$19 mag for Palomar E plates
and R$<$20 mag for UKST R plates). Where the object was detected on both
blue and red plates, the mean stellarness parameter was used. A value of
this parameter $>$1.8 defined an object as a galaxy. The units of this
parameter are Gaussian standard deviations from the the mean stellar
value of zero, so the value of 1.8 is conservative since a few stellar
sources will be included but no galaxies will be excluded.

We initially obtained spectra of galaxy counterparts of 
point-like X-ray sources, whether or not an excess of galaxies was
observed. The first $\approx$15 cases where there was no galaxy excess
were found to be exclusively broad-lined AGN (with FWHM$>$1000 km
s$^{-1}$) or 
low luminosity, normal galaxies. We have thus assumed that the 
X-ray emission from point-like X-ray sources does not arise in intra-cluster
or intra-group gas unless an excess of galaxies is observed at the 
X-ray position, in which case
spectroscopy is required to determine the origin of the X-ray emission.

A large number of telescopes is being used in this work. R band CCD
imaging has been performed at the MDM 1.3m telescope, the Lick 1m Nickel
telescope, the KPNO 0.9m telescope, the CTIO 0.9m telescope, the 
MDM 2.4m telescope and the WIYN 3.5m telescope. Low resolution
spectroscopy has been performed at the KPNO 4m telescope, the CFH 3.6m
telescope, the Lick 3m Shane telescope and the MDM 2.4m telescope. 
Multi-object spectroscopy was used on these telescopes whenever possible.

Finally, we note a possible cause of incompleteness due to the optical
follow-up strategy. Bright, unrelated stars falling in error circles
containing the real, fainter counterparts could mask the true
counterpart. If the X-ray source is extended, we obtain CCD imaging in
any case, and a faint cluster would be visible unless the star was
brighter than R$\sim$15 mag. Only relatively faint (R$\sim$19 mag) stars
are numerous enough at high latitudes ($\sim$0.1 per error circle; Jones
et al.\markcite{j91} 1991) to significantly contaminate the extended
source sample, and stars this faint mask a negligible area of sky. The
small number of point-like X-ray sources with galaxy counterparts which
are cluster candidates (4\% of all point-like X-ray sources) suggests
that masking of point-like X-ray sources by bright stars will also not be
a significant cause of incompleteness.

\subsection{Source identification summary}
The total number of X-ray sources above the limit of 3.5x10$^{-14}$
\ecs (0.5-2 keV) in detected flux in the 86 fields is 283, and 54 of 
these are labelled as
extended. Ten of the point-like sources are also cluster candidates based
on CCD imaging, giving a total of 64 candidates. One large extended
source is faint and of `patchy' appearance, with no excess of galaxies
within the X-ray contours on CCD images. The small number of X-ray
photons in each peak ($<$10) suggests that it is a false source, caused
by a merger of noise peaks and faint point sources, and we have removed
this source from the sample. One other source, which was originally
identified as 3 separate components, each a significant detection but
below the flux limit, has been manually re-inserted in the candidate list
because an excess of galaxies at the position of at least one component
suggested that the source may be a cluster with a large degree of
sub-structure.

In several of the extended sources, inspection of the X-ray contours and
spatial photon distribution clearly shows that they are 2 or 3 close
point sources merged together, and so they have been treated as separate
point sources. Spectroscopy in 2 of these cases has confirmed the
counterparts as AGN. Four extended sources are identified with nearby
individual galaxies which have been resolved with an X-ray extent similar
to the optical extent, and one extended source is identified with a
stellar cluster. We will concentrate on those sources above the flux
limit in total flux, although there are several clusters in the survey
below this limit.

In total, there are 46 candidate clusters and groups of galaxies above
the total flux limit of 6x10$^{-14}$ \ecs (0.5-2 keV), of which five
are coincident with previously catalogued clusters. We have CCD imaging
of all these candidates; in 31 cases there is a clear excess of galaxies
within the X-ray contours. We have spectroscopically confirmed, and
measured redshifts for, 27 of these 31. At least 10 of the remaining 15
cluster candidate error circles contain a spectroscopically confirmed 
broad-lined AGN 
which contributes sufficient flux to put any remaining extended component
below the survey flux limit. We suspect that most of the X-ray emission
in some of the other 5 candidates will not originate in a hot
intra-cluster medium; however, until further spectroscopy is performed,
we label these objects as ``possible'' clusters. We will construct
log(N)-log(S) relations both with and without the ``possible'' clusters.

In four of the confirmed  clusters the 
X-ray contours indicate that a significant level of emission arises
in point sources within the sky area of the cluster, usually from
galaxies within the cluster itself. In these cases an estimate
of the flux from the point sources has been made and the flux
subtracted from the total.  All but one of these clusters are
of low luminosity and at low redshifts z$<$0.3, and thus will
not affect the conclusions based on the high redshift clusters in
our sample.  The individual galaxy luminosities are naturally expected to
be the highest fraction of the intra-cluster medium luminosity
in the lowest luminosity clusters.

\subsection{Estimated redshifts} 
To estimate whether the redshift is above z=0.3 for the minority of 
clusters for which we 
have no spectroscopic measurement, we use the crude approximation that
if the brightest cluster galaxy (BCG) has R$>$18 mag, then 
the cluster has z$>$0.3. This is based on
the Hubble diagram results of Sandage\markcite{s72} (1972) and Hoessel Gunn \&
Thuan\markcite{h80} (1980) and is 
consistent with the clusters for which we do have redshifts.
For  17.5$<$R$_{BCG}<$18.5 we consider the photometric redshift estimate
to be uncertain (partly because of the intrinsic scatter in the Hubble
diagram and partly because of the uncertainty in some of our magnitude
estimates). We show below that our conclusions, based on 
the high-redshift 
counts of clusters, are not sensitive to the 
magnitude chosen to divide the z$>$0.3 and z$<$0.3 samples.

\section{Results}

The integral log(N)-log(S) relation for all 31 optically confirmed
clusters in the WARPS sample is shown in Figure 2, together with the data
from other X-ray selected cluster surveys. Shown on the abscissa is total
flux in the 0.5-2 keV band (all fluxes quoted are for the 0.5-2 keV band
unless explicitly stated otherwise), where ``total flux'' refers to the
flux extrapolated to surface brightnesses below the detection limit. The
WARPS points (shown as solid circles) overlap in flux with the faint end
of the {\it Einstein} EMSS and occupy the gap between the EMSS and the
deep ROSAT survey of Rosati et al.\markcite{r95} (1995). An extrapolation
of the ROSAT BCS counts at bright fluxes (Ebeling et al.\markcite{e97b}
1997b; shown as many small circles), which have a slope of -1.39, is
shown by the dashed line. The WARPS counts lie on this line above a flux
of $\approx$1.5x10$^{-13}$ \ecs, but fall below the extrapolation at
fainter fluxes. The WARPS log(N)-log(S) was constructed using a sky area
calculated separately for each cluster, taking into account its total
flux and its angular core radius. The sky area as a function of these two
parameters is shown in Figure 8 of Paper I. The number density of
confirmed clusters at total fluxes $>$6x10$^{-14}$ \ecs (0.5-2 keV) is
1.8$\pm$0.34 deg$^{-2}$.

The integral log(N)-log(S) relations from the four different surveys
shown in Figure 2 are in reasonable agreement. We investigate the
consistency between the WARPS and the EMSS results below, but first we
note that the general consistency is particularly impressive because each
of the four surveys used independent data (from two different X-ray
missions) and, importantly, independent source detection algorithms. We
thus have some confidence in the completeness of the samples.

A maximum likelihood fit of a power law $N(>S)=KS^{-\alpha}$ deg$^{-2}$
to the WARPS counts (using the method of Murdoch et al.\markcite{m73}
1973, which effectively fits the differential counts) at total fluxes
between 6x10$^{-14}$ and 5x10$^{-13}$ \ecs yields
$\alpha$=0.93$^{+0.36}_{-0.34}$ and K=8.8x10$^{-13}$ when $S$ is measured
in \ecs. A Kolmogorov-Smirnov test confirms that the data are not
significantly different from this power law fit (55\% probability that
the two are different). An extrapolation of the BCS (0.5-2 keV) counts
predicts 3.1 clusters deg$^{-2}$ at the WARPS total flux limit, compared
to the 1.8$\pm$0.34 deg$^{-2}$ observed, and is rejected at a probability
of $<10^{-2}$ (even if all the possible clusters are included), showing
that there is a statistically significant turnover. Although, at the flux
level probed by WARPS, this turnover is largely due to the increased
cosmological stretching of the survey volume, it also reflects the shape
and amplitude of the high redshift cluster XLF.

To measure the slope, we make the simplifying
assumption that the area of sky surveyed is independent of the
angular core radius of the clusters. For the majority of the
clusters (65\%) with core radii between 0.25 arcmin and 0.6 arcmin,
this is accurate to within 7\%. A better method would be to perform
a joint fit to determine the slopes in both flux and core radius. 
However, the
effect of varying the value of the assumed constant core radius
between 0.35 arcmin and 0.55 arcmin is to vary the measured 
slope between 0.93 and 0.97,
a small variation given the statistical errors caused by the
small numbers of clusters in the sample.

The short dashed line just above the WARPS points in Figure 2 indicates the 
log(N)-log(S) relation obtained when all the ``possible'' clusters
are included. The change when these objects are included is small;
there is an increase at the faint limit
equal in size to the error bar just visible on the faintest WARPS point.
A small constant additive 
correction of 0.04 deg$^{-2}$ has been added to the WARPS 
integral log(N)-log(S) points to correct for the bright clusters
at fluxes $>$1x10$^{-12}$ \ecs that did not appear in the
survey because the area of sky sampled was too small. The value of
0.04 deg$^{-2}$ corresponds to a flux of 1.4x10$^{-12}$ \ecs 
in the BCS log(N)-log(S) relation.

The BCS data of Figure 2 are taken directly from Ebeling et al.
\markcite{e97b} (1997b).
The EMSS data are also taken from Ebeling et al.\markcite{e97b} (1997b),
who derived the EMSS counts using the appropriate sky coverage and
correction to total flux. We follow Henry et al.\markcite{h92} (1992) and
assume a constant core radius of 250 kpc for the EMSS clusters. We have
corrected the EMSS counts from the {\it Einstein} 0.3-3.5 keV band to the
ROSAT 0.5-2 keV band using a constant factor of 1.7, appropriate for a
Raymond and Smith
\markcite{r76} (1977) thermal spectrum of temperature 4 keV and abundances of
between 0.25 and 1 times cosmic abundance.
We note that this approximation gives results accurate to $\lesssim$ 5\%
when applied to the BCS log(N)-log(S) of Ebeling et al.\markcite{e97b} 
(1997b) derived in the 0.3-3.5 keV band correctly, using an
individual temperature for each cluster.

Although the integral log(N)-log(S) relation of Figure 2 gives a good
overview, detailed comparisons can be misleading because
the data points within each survey are not statistically
independent. In order to
comment on, for example, the completeness of the EMSS in the light of the
WARPS results, we turn to the differential log(N)-log(S) relation of
Figure 3 in which the error bars and the data points within each 
survey are all statistically independent.
Figure 3 contains the same data as Figure 2.
The EMSS points lie below the WARPS points but they are not significantly
different (\chis=2.54 for 2 degrees of freedom (dof),
corresponding to 28\% 
probability that they arise from the same distribution). 
The maximum WARPS cluster counts produced by including the 
``possible'' clusters is shown by the short dashed line.

The number of WARPS clusters at redshifts z$>$0.3 and above the total
flux completeness limit is 12 (there are an additional 3 clusters below
this flux limit which we do not consider further). Of these 12, 10 have
measured redshifts, and 2 have redshifts estimated to be above z=0.3 from
the magnitude of the brightest galaxy. The differential log(N)-log(S)
relation of the high redshift clusters is shown in Figure 4. There is
good agreement between the WARPS counts and the EMSS counts. The maximum
and minimum WARPS log(N)-log(S) relations are shown as dashed lines in
Figure 4. The maximum number of z$>$0.3 WARPS clusters is 18, if the
``possible'' clusters are included and the brightest galaxy magnitude
corresponding to z=0.3 is assumed to be R$_{BCG}$=17.5 mag instead of
R$_{BCG}$=18 mag. The minimum number of z$>$0.3 clusters is 10, if
R$_{BCG}$ at z=0.3 is taken to be 18.5 mag and we remove the 2 fields
where the observation target was a high redshift cluster and which
contained other high redshift clusters (although at very different
redshifts from the targets).

The high redshift log(N)-log(S) relation
is more sensitive to evolution than the log(N)-log(S) relation 
of clusters at all redshifts, and it is from the high redshift data
that we will draw our conclusions about the rate of evolution 
of low luminosity clusters. First, though, we describe the models which we
use to predict the number of clusters.

\section{Predicted counts and models of cluster evolution}

In order to predict the expected number of clusters as a function of flux
and redshift, we first integrate the zero redshift XLF
assuming no evolution of the XLF with redshift,
but including K-corrections and the effect of the co-moving volume element
for the assumed value of q$_{\circ}$. We use the BCS zero redshift
XLF of Ebeling et al.\markcite{e97a} (1997a). The details of the K-correction,
which is in general a 10\%-20\% effect, are given
in an Appendix. 
We then investigate the effect of pure
density evolution on the predicted log(N)-log(S) relation, and lastly compare
the predictions of the more physically motivated
evolution models of Mathiesen \& Evrard\markcite{m97} (1997)
with the data. All the models assume that all the X-ray flux 
within a given energy band from each cluster has been detected and that
all the observational detection limitations have been removed. This is not
quite true. The detected fluxes have been converted to total
fluxes and corrections have been made (via the sky area surveyed)
for the slightly lower detection probability 
of {\it detected} sources of large angular size compared with those of
small size (for 
a given total flux).
Sources will be missing from the survey if they are of such a large angular 
size that all their flux falls below
our surface brightness threshold, even though the total flux is above the
survey limit. This incompleteness, including the effect of cosmological 
surface brightness dimming, is estimated in Section 5.4 and found to be
small.

\subsection{Clusters at all redshifts}

The two smooth curves in Figure 2 show the predicted log(N)-log(S) relation for
all clusters in the survey assuming no evolution of the XLF with redshift; 
the integration of the Ebeling et al.\markcite{e97a} 
(1997a) 0.5-2 keV XLF was performed over the redshift range 0$<$z$<$2 and the 
luminosity
range 1x10$^{42}$ erg s$^{-1}$ $<$L$_{X}<$ 1x10$^{47}$ erg s$^{-1}$,
encompassing all detected cluster luminosities.
At the 
flux limit of the WARPS survey there is little difference between the 
predictions for \qo=0.5 (lower curve) and \qo=0 (upper curve). Both curves
fit the WARPS data and the Rosati et al. \markcite{r95} (1995) data well. 
For \qo=0 our use of the BCS XLF is not strictly valid, since the  
BCS XLF was derived assuming
\qo=0.5. However, since the median BCS cluster redshift is z$\approx$0.1,
the effects
of the assumed value of \qo ~on the value of the BCS XLF will be 
small. 

In Figure 3 we quantify the similarity between a \qo=0.5 no-evolution
model (shown as a solid line) and the observed differential log(N)-log(S)
relations. The WARPS data are consistent with a no-evolution model, both
including and excluding the ``possible'' clusters. The EMSS data lie
slightly below the no-evolution model. Assuming only Poisson errors, the
\chis ~for EMSS clusters at fluxes $>$10$^{-11}$ \ecs is 19.2 (for 8
dof), corresponding to 1\% probability that the data are consistent with
the no-evolution model. However, a systematic increase in flux by a
factor of 1.25 in the EMSS data would make them consistent with the model
(at 44\% probability). A systematic error of that size is very possible,
given the mean EMSS conversion factor from detected to total flux of 2.5
and the assumed constant core radius of 250 kpc (in contrast to WARPS
where the core radius is estimated for each cluster independently, and
the mean conversion factor from detected to total flux is 1.4). There is
an additional small uncertainty in the conversion from the EMSS 0.3-3.5
keV band to the 0.5-2 keV band. Ebeling et al.\markcite{e97b} (1997b)
show that the difference in the EMSS counts introduced by assuming a
constant core radius of 300 kpc instead of 250 kpc is a factor which
varies with flux between values of $\approx$1.05 and 1.2, almost
sufficient to account for the observed difference. We investigate in
detail possible systematic differences between flux measurement methods
in Appendix A, using ROSAT PSPC data of EMSS clusters, and find that the
EMSS fluxes may be too small by a factor of $\approx$1.2-1.3. Thus we
conclude that the WARPS and EMSS all-redshift cluster log(N)-log(S)
relations are in good agreement, especially if this correction is
applied, and that they show no evidence for evolution of the XLF.

This is not inconsistent with the result of Henry et al.\markcite{h92}
(1992), who found evidence for evolution at high redshifts in the EMSS
data, since here we are not including any redshift information and the
log(N)-log(S) relation blurs the differences between low and high
redshifts. In the next section we examine the high redshift log(N)-log(S)
relation separately in order to clarify the situation.

\subsection{Clusters at high redshifts}
In Figure 4 we show the predicted differential log(N)-log(S) relation for
clusters at z$>$0.3 assuming the same zero redshift XLF and integration limits
as above (taking into account the lower limit imposed on luminosities
by the z$>$0.3 redshift limit, which is 2.6x10$^{43}$ erg s$^{-1}$ at the 
WARPS flux limit).
 The solid line is for \qo=0.5 and no-evolution; this is a
good match to the WARPS data (which are dominated by the faintest bin).
This is true for the range of
log(N)-log(S) relations both including and excluding the possible
clusters, 
as shown by the dashed lines. The EMSS data, however, fall
systematically below the no-evolution prediction. Assuming Poisson
statistics alone, the EMSS data are inconsistent with the no-evolution
prediction (\chis=20 for 4 dof or $<$0.1\%
probability) but a systematic flux increase by a factor of 1.25 in the 
EMSS data would 
remove the inconsistency (\chis=3.0 or 56\% probability).

In order to quantify the level of evolution allowed by the data, we have
predicted  log(N)-log(S) relations assuming pure density evolution of the
XLF $\phi$(z) of the form 

$\phi$(z) = $\phi(0) (1+z)^{\alpha_D}$

which is applied equally to all luminosities. This simple parameterisation 
provides a convenient description of the data for comparison with e.g. detailed
hydrodynamic or N-body models of cluster evolution.
The dashed lines in Figure 4 were calculated using
$\alpha_D$=-2 and $\alpha_D$=-3. The $\alpha_D$=-2
parameterisation is consistent with the EMSS data, but $\alpha_D\le$-3
is inconsistent with the WARPS data (at $<$1\% probability), and 
$\alpha_D=$-2 is only marginally inconsistent with the WARPS data 
(2\% probability).

Although the z$>$0.3 log(N)-log(S) relation of Figure 4 does not
show evidence of inconsistency with the \qo=0.5 no-evolution prediction (given
a possible EMSS systematic error), there is a trend in which the lowest
(WARPS) flux point lies just above the prediction whereas the brightest (EMSS) 
flux points lie below the prediction, even if their flux is increased 
systematically by a factor of 1.25. We will thus check for differences
 between the WARPS and EMSS samples.
One difference is the redshift 
distribution at z$>$0.3. However,
because the WARPS sample has a fainter limiting flux than the EMSS
sample, it will have a higher mean redshift, and thus should show
more evolution, not less, assuming any evolution is a monotonic 
function of redshift.

The more important difference between the WARPS and EMSS
high redshift clusters is the range of X-ray luminosities covered by the
two samples.
The luminosities of the high redshift WARPS clusters lie in the range 
from 4x10$^{43}$ h$_{50}^{-2}$ erg s$^{-1}$ to 
2x10$^{44}$ h$_{50}^{-2}$ erg s$^{-1}$ (0.5-2 keV, q$_{\circ}$=0.5), 
whereas the EMSS clusters lie in the range from 1x10$^{44}$ h$_{50}^{-2}$ erg
s$^{-1}$ to 1.5x10$^{45}$
h$_{50}^{-2}$ erg s$^{-1}$ (0.5-2 keV, q$_{\circ}$=0.5). We will
investigate whether the evolution
rate is luminosity-dependent.

\subsection{Different evolution at low and high luminosities}

In Figures 5 and 6 we show the log(N)-log(S) relation for a restricted subset
of clusters; high redshift (z$>$0.3), low luminosity clusters (Figure 5)
and high redshift (z$>$0.3), high luminosity clusters (Figure 6). In both
Figures the no-evolution prediction is shown by a solid line.
The low luminosity clusters of Figure 5 include all the WARPS clusters
at z$>$0.3 and 10 EMSS clusters with L$_X<$3x10$^{44}$ h$_{50}^{-2}$
erg s$^{-1}$ 
(0.5-2 keV,  q$_{\circ}$=0.5) where a conversion factor of 1.7 between
the Einstein 0.3-3.5 keV and ROSAT 0.5-2 keV bands has been used (see 
Section 4).
The EMSS and WARPS counts are in good agreement. Although there are
large errors on both datasets, they are both 
consistent with the no-evolution
prediction (\chis=2.34 for 2 dof, corresponding to 31\% 
probability for the WARPS points, and \chis=2.25 for 3 dof,
corresponding to 52\% probability for the EMSS points, which
have an even higher probability if their flux is increased by a factor of
1.25). As before, the  prediction is based on the
BCS zero redshift XLF which was integrated over 0.3$<$z$<$2 and 
10$^{42}<$L$_X<$3x10$^{44}$ erg s$^{-1}$.

In contrast, the high luminosity clusters shown in Figure 6 fall a factor
$\approx$2.5-3 below the no-evolution prediction, which was obtained by
integrating the zero redshift BCS XLF over 0.3$<$z$<$2 and
3x10$^{44}<$L$_X<$10$^{47}$ erg s$^{-1}$. There are no WARPS clusters
with luminosities this high, so this Figure contains only data from the
EMSS, for which \chis=28.2 (for 4 dof), corresponding to a probability of
$<$10$^{-4}$ that the data and the no-evolution prediction are consistent
(assuming the error due to the small number of EMSS clusters dominates
the error in the prediction). The probability is still only 1\%
(\chis=13.1) if a systematic flux increase of 1.25 is applied to the EMSS
data. The negative evolution EMSS result of Henry et al.\markcite{h92}
(1992), and the comparison of the EMSS and BCS luminosity functions of
Ebeling et al.\markcite{e97a} (1997a), is confirmed.

We parameterise evolution using the pure density evolution index
$\alpha_D$ as before. The long dashed lines in Figures 5 and 6 show
the predictions for various values of $\alpha_D$, both positive
and negative. The number of WARPS clusters observed at z$>$0.3,
with total flux $>$6x10$^{-14}$ \ecs (0.5-2 keV) and of low
luminosity L$_X<$3x10$^{44}$ h$_{50}^{-2}$ erg s$^{-1}$ is 0.73$\pm$0.34
deg$^{-2}$ (at 90\% confidence), compared to 
the no-evolution prediction of 0.63 deg$^{-2}$ and
corresponding to -1.2$<\alpha_D<$+1.8 (at 90\% confidence).
At the same redshift limit and the higher EMSS flux
limit of 1.3x10$^{-13}$ \ecs (0.5-2 keV), the number of high luminosity EMSS
clusters (L$_X>$3x10$^{44}$ h$_{50}^{-2}$ erg s$^{-1}$) observed is
0.053$\pm$0.021 deg$^{-2}$ (at 90\% confidence), compared
to the no-evolution prediction of 0.16 deg$^{-2}$ and corresponding to
-3.5$<\alpha_D<$-1.5 (at 90\% confidence), a significantly different 
range of $\alpha_D$. If the EMSS flux limit is actually a factor of 1.25
higher, the no-evolution prediction becomes 0.13  deg$^{-2}$,
corresponding to -3$<\alpha_D<$-1.3.

The short dashed lines in Figure 5 show the possible range of the
WARPS counts given the uncertainties due to the as yet unidentified cluster
candidates, the clusters with estimated redshifts, and also include
the effect of removing the two fields which had high redshift 
cluster targets. The effects of all these uncertainties is 
of similar size as the statistical error.
The lowest possible WARPS counts are still consistent
with no-evolution but inconsistent with $\alpha_D$=-3. The highest 
possible WARPS counts may be more consistent with weak positive
evolution than no evolution, but we cannot distinguish 
between these possibilities at this stage. We are in the process of
expanding the sample size in order to investigate this possibility.

So far, we have only considered density evolution. An alternative is pure
luminosity evolution in which the XLF scales only in luminosity with
redshift, such that $L^*(z)=L^*(0)(1+z)^{\alpha_L}$ where $L^*$ is the
characteristic luminosity of the Schechter function XLF. Because the XLF
is steepest at high luminosities, a single value of $\alpha_L\approx$-1
fits both the low luminosity and high luminosity high redshift
log(N)-log(S) relations of Figs 5 \& 6. Although this parameterisation is
attractive because of its simplicity (it is independent of luminosity),
pure luminosity evolution is not consistent with the high redshift
luminosity function of Henry et al. \markcite{h92} (1992) when compared
by Henry et al. with their lower redshift luminosity functions or when
compared with the more accurately measured low redshift BCS luminosity
function by Ebeling et al.\markcite{e97a} (1997a).

\subsection{Surface brightness dimming}

The models described above assume that all the clusters above a given
flux limit are detected and that the detected flux is corrected to a
total flux. They omit surface brightness dimming effects which in
principle could cause a cluster of large angular size to be missed
completely from the survey. In this Section we estimate that the number
of clusters missed because they fall completely below the surface
brightness limit is a small fraction of the total.

We adopt a simple empirical approach. In order to predict the cluster
angular core
radius-flux distribution and compare it with the survey sensitivity, we
need to assume a core radius-luminosity relation. Based on the
virial theorem and simple 
scaling arguments (eg Kitayama \& Suto\markcite{k96} 1996) we adopt

\[
r_c={ {250}\over{h_{50}} } \left({ {L_{44}}\over{5} }\right)^{0.2}  ~~~~~~~kpc
\]

where we have normalised the core radius $r_c$ to be 250 h$_{50}^{-1}$ kpc
for a cluster  of luminosity $L_{44}=5$ in units of 10$^{44}$ erg s$^{-1}$.
We make the simplifying assumption that $\beta={2\over3}$ for all clusters.
This relation is in reasonable agreement with the measurements of 
nearby clusters of 
Jones \& Forman\markcite{j84} (1984) and Kriss et al.\markcite{k83} (1983). 
The mean Jones \& Forman values of $r_c$ for clusters with centrally
dominant galaxies (`XD' clusters) agree with the above relation to within
25\% for cluster luminosities of 10$^{43}$ erg s$^{-1}$ to 10$^{45}$ erg
s$^{-1}$ and for $\beta$ fixed at 0.6. 
For groups with luminosities $<$10$^{43}$ erg s$^{-1}$ the
above relation is not such a good description, although the
general trend is correct, and there is a large
scatter in the observed core radii of local groups (eg Mulchaey et al. 
\markcite{m96} 1996). 

Given the above relation, we integrate the BCS XLF of Ebeling et
al.\markcite{e97a} (1997a) over 0$<$z$<$2 and 10$^{42}<L_X<10^{47}$ erg
s$^{-1}$ as described in Section 5 to obtain the predicted flux-angular
core radius distribution. The clusters are predicted to occupy a region
of the flux-angular core radius plane to which WARPS has good
sensitivity, as measured by the simulations described in Paper I. At the
lowest fluxes (6x10$^{-14}$ to 8x10$^{-14}$ \ecs) and large core radii
($>$0.6 arcmin) the detection probability is $<$80\%. The fraction of
clusters predicted within this flux range with core radii $>$0.6 arcmin
is only 3\% of the total number of clusters in the flux range. Virtually
no clusters ($<$1\%) are predicted at core radii $>$0.8 arcmin within
this flux range. This corresponds to a luminosity of 8x10$^{43}$
h$_{50}^{-2}$ erg s$^{-1}$ and a core radius of 340 h$_{50}^{-1}$ kpc at
a redshift of z=0.5 (for q$_0$=0.5). We would only be able to detect such
objects in 40\% of the survey area. At further extremes a cluster of core
radius 1.2 arcmin (or 510 h$_{50}^{-1}$ kpc at z=0.5) at the flux limit
would only be detectable in 10\% of the survey area. Jones \& Forman
(1984) found that 4 out of 30 (13\%) Abell clusters at z$<$0.06 and
$L_X<10^{44}$ erg s$^{-1}$ had core radii $>$350 kpc, for $\beta$=0.6,
and all of these were `nXD' systems without centrally dominant galaxies.
Only 2 out of 30 (7\%) had core radii $>$500 kpc.

In general, clusters are predicted to mostly populate regions of the
flux-angular core radius plane where the detection probability is high,
at least in the WARPS survey. Of course, scatter in the $r_c-L_X$
relation and the inclusion of the less common clusters without centrally
dominant galaxies will result in some clusters being lost from the
survey, but we expect that the number lost in any flux range will be
$\lesssim$10\% of the total, particularly at z$>$0.3 where the cluster
luminosities are always $>$10$^{43}$ erg s$^{-1}$, a luminosity range
where the sizes have been well sampled, at least in the local Universe.

\subsection{Comparison of the WARPS and RIXOS number count results}
The RIXOS cluster survey of Castander et al.\markcite{c95} (1995) found
strong evidence for negative evolution at redshifts z$>$0.3 from a ROSAT
survey to a similar flux limit and covering a similar area of sky as that
used here. Since the conclusions of Castander et al.\markcite{c95} are
quite different to ours, we investigate here possible reasons for the
discrepancy.

Firstly we compare directly the surface density of z$>$0.3 clusters, not
including the instrumental effect of varying sensitivity across
the PSPC field of view, as these are approximately
 the same for both surveys. RIXOS has 
5 clusters at z$>$0.3 from 14.9 deg$^{2}$ or 0.33$\pm$0.15 deg$^{-2}$ above 
a detected flux
of 3.0x10$^{-14}$ erg cm$^{-2}$ s$^{-1}$ (0.5-2 keV). This flux 
limit is close to,
but slightly less than, our limit of 3.5x10$^{-14}$ in detected flux,
and so if anything the RIXOS survey should measure a higher surface
density of clusters. However, we find 14 z$>$0.3 clusters from
16.2 deg$^{2}$ or 0.86$\pm$0.22 deg$^{-2}$, 2.5 times the 
RIXOS density and significantly different at
the 95\% level.

In this paper we have taken the approach of correcting the measured
source fluxes to obtain an estimate of the total flux from each cluster.
Castander et al. \markcite{c95} take a different approach by assuming all
clusters have the same core radius, modeling the detection of the
clusters and including the detection efficiency in the n(z) predictions
obtained by integrating different evolutionary XLFs.  Although detailed
comparisons between the two approaches are difficult to make, a
comparison of the number of clusters detected relative to the prediction
of the no evolution model in each case should take into account the
differences. In the RIXOS survey, at z$>$0.3 the number detected (5) is
0.28 times the number predicted (18) from Figure 1 of Castander et
al.\markcite{c95} In WARPS the number detected (in the range of total
flux where WARPS is complete at z$>$0.3, i.e. 6x10$^{-14}$ \ecs -
2x10$^{-13}$ \ecs) ~is at least 12, or 0.89 times the number predicted
from the no-evolution model (for q$_{\circ}$=0.5 as used by Castander et
al.\markcite{c95}). This is 3.2 times the number of clusters observed in
the RIXOS survey (in each case relative to the no-evolution model), and
this difference leads to the different conclusions in this paper and
those of Castander et al.\markcite{c95}

One might think that the difference could be due to differences in the
optical follow-up strategies. Castander et al.\markcite{c95}
spectroscopically identified nearly all (95\%) of the detected X-ray
sources.  Source classification here is based partly on X-ray extent and
partly on optical imaging, so any difference due to the follow-up
strategies would result in fewer clusters detections in WARPS, not more.

Another, more compelling, hypothesis is that the discrepancy between the
WARPS and RIXOS results is due to fundamental differences in the X-ray
source detection algorithms used in the two surveys. The RIXOS source
detections were based partly on an algorithm optimised for point sources.
As shown in Paper I, a point-source based algorithm will severely
underestimate the flux from extended sources in the PSPC data. A
systematic flux underestimate of 30\% (less than the typical correction
we apply for undetected flux below the surface brightness threshold) will
also reduce the number of clusters by $\approx$30\% at the WARPS flux
limit, partly explaining the discrepancy. If this is the case, RIXOS
completeness might at first sight be expected to increase with redshift,
since clusters of the same linear size will have a smaller angular size
at higher redshift, and suffer less flux loss. However, at higher
redshift, the constant survey flux limit means that clusters of higher
luminosity and thus larger linear size will be observed, and the net
result is that the mean angular size increases only slightly over the
flux range where most clusters are detected ($\sim$6-20 x10$^{-14}$ \ecs;
see fig 8 of Paper I). Thus the RIXOS incompleteness should not be a
strong function of redshift, at least at z$>$0.2. Of course the use of a
point source detection algorithm does not only entail the risk of
systematically underestimating the fluxes of extended sources; the latter
may be missed altogether even at intermediate redshifts.

\section{Discussion}

We have measured a luminosity-dependent rate of evolution for clusters of
galaxies over the redshift range z$\approx$0.3-0.4 to z=0.
Our result that the cluster X-ray luminosity function  
does not evolve at low luminosities (or at least evolves less 
negatively than at high 
luminosities) supports the hierarchical model of the growth of structure,
in which less massive clusters require less time to form.
The predictions of CDM using the Press-Schechter\markcite{p74} (1974) 
formalism as plotted
in e.g. Efstathiou \& Rees\markcite{ef88} (1988) and Peacock
\markcite{p91} (1991) are
that the number density of objects with the mass of X-ray luminous
clusters evolves strongly over the range z=0 to z=1, whereas less massive
objects are predicted to have less evolution. Observationally, our result
is in agreement with the recent survey of Collins et al.\markcite{c97}
(1997) but disagrees with Castander et al.\markcite{c95} (1995). There
was also tentative evidence in the EMSS luminosity functions of Henry et
al.\markcite{h92} (1992) and the comparison of the EMSS and BCS
luminosity functions in Ebeling et al.\markcite{e97a} (1997a) for
different levels of evolution at different luminosities.

We have compared this initial dataset with a 
luminosity-dependent density evolution parameterisation. Realistically, a
combination of both luminosity and density evolution is expected. The evolution
of the hot gas density and hot gas mass will largely determine the luminosity
evolution, while the number of clusters of a given mass in a hierarchical model
depends on the rate of 
formation (from the merging of smaller clusters) and the rate of 
destruction (by merging into larger clusters). The combination of these
effects determines the overall evolution of
the XLF, which is thus dependent on the energetics of galaxy evolution 
(including the level of
heat input into and cooling of the ICM) and on the cosmology and 
the primordial fluctuation spectrum.

\subsection{Cosmological models and Thermal Histories}

One of the variants of the cold dark matter (CDM) model of structure
growth (designed to match the observed level of galaxy clustering on
large scales together with the COBE results) is the cold+hot dark matter
model (CHDM). For $\Omega_0=1$ ($\Omega_{hot}$=0.3, $\Omega_{cold}$=0.6,
$\Omega_{baryon}$=0.1) Bryan et al.\markcite{b94} (1994) used a
hydrodynamic plus N-body model to predict strong negative evolution of
the 2-10 keV cluster XLF at z=0.5 and $L_X<10^{44}$ erg s$^{-1}$. If we
assume a similar level of evolution is predicted in the 0.5-2 keV band,
then Bryan et al.\markcite{b94} predict $\alpha_D<-3$ in the
parameterisation used here. Jing \& Fang\markcite{j94} (1994) also
predicted strong negative evolution of the cluster temperature function
in the CHDM model. The observations presented here rule out these CHDM
models, if taken at face value.

Furthermore, models in which the cluster gas simply scales as the mass
distribution (e.g. Kaiser\markcite{k86} 1986) have difficulty in
simultaneously reproducing the observed temperature function (Henry \&
Arnaud\markcite{h91} 1991) and XLF properties.  This latter problem can
be helped by assuming that the central gas entropy is largely due to heat
input at an early epoch (e.g. Evrard\markcite{ev90} 1990,
Kaiser\markcite{k91} 1991) after which the gas settles adiabatically into
the dark matter potential wells and is little heated by subsequent merger
shocks. The X-ray gas distribution within the cluster is then more
dependent on the total cluster potential than the density profile of the
dark matter (which increases as $(1+z)^3$) and the cluster XLF is
expected to show some negative evolution. Independently, recent
observations of cluster gas metallicities (e.g. Loewenstein
\& Mushotzky\markcite{l96} 1996) have provided good evidence that 
the widely distributed
metals were produced by type II supernovae with an energy budget
sufficient to provide significant heating at some epoch $z\gtrsim 2$.
Measurements of the cluster $L_X-T$ relationship at $z=0.4$ (Mushotzky
\& Scharf\markcite{ms97} 1997) for luminous clusters ($L_{bol}\simeq 3\times 
10^{45}$ erg
s$^{-1}$) show no evidence for evolution from $z=0$, and no evolution of
the temperature function has been found by Henry\markcite{he97} (1997) up
to $z=0.33$, in accord with a preheating scenario.

One result of the Kaiser\markcite{k91} (1991) preheating model is that
much weaker negative evolution of the XLF is expected for clusters of
luminosity below 10$^{45}$ erg s$^{-1}$. This prediction is supported by
Bower\markcite{b97} (1997) whose `constant entropy' model for n=-1 and
$\Omega_0$=1 actually predicts mild positive evolution of the
differential XLF for $L_X\approx10^{44}$ erg s$^{-1}$, but negative
evolution by a factor $\approx$3 for $L_X\approx$4x10$^{44}-10^{45}$ erg
s$^{-1}$, as observed. This prediction is in good, qualitative, agreement
with the results presented here which therefore provide the first
confirming evidence for this type of thermal history, based on cluster
population statistics alone.

Mathiesen \& Evrard\markcite{m97} (1997) have used the WARPS log(N)-log(S) data
of all clusters, as presented here, together with the updated data of
Rosati et al.\markcite{r95} (1995), to constrain the parameters of
a semi-analytical model based on a total mass to X-ray luminosity
relation of the form:

\[
L_X=L_{15}M^p(1+z)^s
\]

Mathiesen \& Evrard use the Press-Schechter\markcite{p74} (1974)
formalism to describe the rate of growth of dark matter halos, which
includes merging as larger halos grow faster than nearby smaller halos
and `swallow' them. The Press-Schechter mass function is converted to a
luminosity function using the above relation, and the parameters $L_{15}$
and $p$ are determined by fitting to the local XLF of Ebeling et
al.\markcite{e97a} (1997a). The parameter $s$ describes the evolution of
the luminosity-mass relation in co-moving coordinates and includes the
combined effects of cooling of the ICM via the expansion of the Universe,
together with any heating of the ICM from galaxy winds or cooling via
cooling flows. A value of $s$ of $\approx$3 is indicative of constant
entropy of the ICM in the cluster core with redshift (Evrard \&
Henry\markcite{ev91} 1991, Bower
\markcite{b97} 1997). Two models
which give good fits to the log(N)-log(S) data are shown in Figure 3.
Model (a) has $\Omega_0$=1, n=-1, s=6 and no cosmological constant.
Model (b) has $\Omega_0$=0.3, n=-1, s=2 and again no cosmological constant.
Both models, although containing the evolution inherent in the Press-Schechter
formalism and the evolution of the above luminosity-mass relation,
give similar log(N)-log(S) predictions as a simple model in
which the XLF does not evolve.
Some of the evolutionary terms evidently work in opposite directions, 
partly cancelling each other.

Mathiesen \& Evrard find that $\Omega_0$ and $s$ are constrained
by the log(N)-log(S) data
such that $\Omega_0$=1 requires $s\ge3$ and $\Omega_0<0.2$ requires
$s<$2.5, and that these conclusions are relatively insensitive to the
value of n and the presence of a cosmological constant in a flat 
Universe. So 
for $\Omega_0$=1, preheating of the X-ray gas to provide the
initial cluster core entropy (and possibly further heating via
cluster merger shocks or galaxy winds) is probably required, as found above.
Less theoretical modelling has been performed for low $\Omega_0$, 
and it is difficult to comment in detail on whether 
models including preheating of the X-ray gas are preferred. The
results of Mathieseon \& Evrard\markcite{m97} (1997) suggest that
cooling mechanisms may be dominant if $\Omega_0<0.2$.

\subsection{Predictions for future surveys}
Figure 6 reinforces the rareness of high redshift, high luminosity
clusters. Because the log(N)-log(S) relation of these clusters is so
flat, surveys which probe to faint fluxes over even a relatively large
area of sky are not an efficient way of finding them. A serendipitous
survey at faint fluxes $\sim10^{-14}$ \ecs (eg XMM pointed observations)
would need to cover 200 deg$^2$ in order to detect $\sim$30 high redshift
(z$>$0.3), high luminosity ($>$3x10$^{44}$ erg s$^{-1}$) clusters if the
negative evolution observed at z$\approx$0.33 ($\alpha_D=-2$) continues
to higher redshifts. These clusters would represent only 0.15\% of all
the X-ray sources. However, such a survey would detect $\sim$1000 high
redshift, low luminosity clusters, assuming no evolution at low
luminosities. A more efficient way of finding high redshift, high
luminosity clusters would be a large area (e.g. 5000 deg$^{2}$) survey at
a relatively bright flux limit (e.g. 2x10$^{-13}$ \ecs) with sufficient
spatial resolution ($<$20 arcsec) to resolve high redshift clusters (e.g.
an XMM slew survey, or the ROSAT \& ABRIXSAS All Sky Surveys if the
spatial resolutions are adequate). Such a survey would provide an order
of magnitude increase in the number of X-ray selected high redshift, high
luminosity clusters: $\approx$200 clusters at z$>$0.3 and
L$_X>$3x10$^{44}$ erg s$^{-1}$ (of which $\approx$35 would be at z$>$0.7)
again assuming negative evolution continues to higher redshifts. They
would represent 4\% of all the X-ray sources.

\section{Conclusions}
We have presented initial results from an X-ray selected, flux and
surface brightness limited, complete survey of clusters of galaxies at
relatively faint X-ray fluxes. The log(N)-log(S) relation of the clusters
is consistent with previous measurements at both brighter and fainter
fluxes. We have obtained redshifts for most, but not all, of the
candidate clusters above our limit in total flux of 6x10$^{-14}$ \ecs
~0.5-2 keV, including 10 at z$>$0.3 and a further 2 with estimated
redshifts of z$>$0.3. Based on the properties of nearby clusters and our
surface brightness limit, we estimate that few clusters are missing from
our survey, particularly at high redshifts. The X-ray luminosities of the
high redshift clusters lie in the range 4x10$^{43}$ \hes to 2x10$^{44}$
\hes, the luminosities of poor clusters. The number of high redshift, low
luminosity clusters is consistent with no evolution of the X-ray
luminosity function between redshifts of z$\approx$0.4 and z=0. Mild
positive evolution at the faintest luminosities cannot be ruled out. A
limit of a factor of $<$1.7 (at 90\% confidence) is placed on the
amplitude of any pure negative density evolution of clusters of these
luminosities.  An alternative parameterisation is the density evolution
index $\alpha_D$ of the XLF which is constrained to be
-1.2$<\alpha_D<$+1.8 (at 90\% confidence) for low luminosities. This can
be contrasted with the value of -3.5$<\alpha_D<$-1.3 (at 90\% confidence)
for EMSS clusters at similar redshifts but higher luminosities
($>$3x10$^{44}$ \hes).

In a simple interpretation,
this difference in the evolution of the cluster XLF at low luminosities and at
high luminosities 
supports the hierarchical model of the growth of
structure in the Universe. When compared with detailed
modelling, as performed by Kaiser\markcite{k91} (1991), Evrard \& Henry
(1991), Bower 
\markcite{b97} (1997) and Mathiesen 
\& Evrard\markcite{m97} (1997), this evolutionary pattern is matched by models
in which the X-ray gas is preheated at some early epoch, at least
for $\Omega_0$=1. 

We suspect that the higher number of high redshift clusters found in this
survey compared to that of Castander et al.\markcite{c95} (1995) is due
to the higher sensitivity to low surface brightness X-ray emission of the
source detection algorithm used here. Finally, we have investigated
differences in the flux measurement methods used here, in the EMSS, and
by Nichol et al. (1997). We find that the EMSS fluxes may have been
underestimated by 20-30\%, but that the EMSS sample still shows evidence
of negative evolution at high luminosities. The WARPS and EMSS
Log(N)-Log(S) relations for all clusters, while not inconsistent, are in
better agreement if the EMSS fluxes are increased by this amount.

\acknowledgements
This project has benefitted from the help of many people.
We thank Mike Irwin for APM data, 
Geraint Lewis, Lance Miller and Mike Read
for obtaining INT identifications, Greg Wirth for last minute help 
with masks at
Lick, Rem Stone for help at the Lick 40 inch telescope,
the KPNO TAC and the staff
at Kitt Peak, and Richard Mushotzky for stimulating discussions. 
We thank Ben Mathiesen 
\& August Evrard for sharing and discussing their model results,
and the referee, Alastair Edge, for useful comments.
This research has made use of data obtained through the High Energy
Astrophysics Science Archive Research Center Online Service,
provided by the NASA/Goddard Space Flight Center, and through the
STScI  Digitized Sky Survey archive.
Part of this work
was performed while CAS and LRJ were supported at NASA/GSFC by Regular 
and Senior NRC Research Associateships respectively, and ESP was
supported  at NASA/GSFC by a USRA Visiting Scientist Fellowship.
LRJ acknowledges support from the UK PPARC.
HE acknowledges financial support from SAO contract SV4-64008.

\appendix 
\section{Systematic differences in flux measurement methods}

In order to investigate any possible systematic difference in the
methods used to measure the cluster fluxes here and in the EMSS,
we have measured the fluxes of the 14 EMSS clusters
where the X-ray emission is fully contained
within 18 arcmin of the centre of a ROSAT PSPC field (this is
the area we analyze with VTP). We also compare our flux values
with those obtained by Nichol et al (1997) who analyzed the same
ROSAT data but used a different method to measure the fluxes.

We use two methods to measure the EMSS cluster fluxes. The first method
is our standard VTP analysis as applied to all the PSPC fields within
WARPS, including exposure maps in each of the 0.5-0.9 keV and 0.9-2 keV
bands. When analyzing the VTP results, we select a threshold for each
EMSS cluster as we did for the WARPS sources, and apply our standard
correction from detected to total count rate using the estimated core
radius of each cluster. To convert from count rate to flux outside our
Galaxy, we use the column density appropriate for each cluster from
Dickey \& Lockman \markcite{d90} 
(1990), a metallicity of 0.3 and measured temperatures
where available, or the cluster X-ray luminosity-temperature relation
within an iterative procedure to estimate the temperature. The
temperatures range from 2.8 keV (estimated) to 10.2 keV (measured). We
also use these temperatures and column densities to convert each cluster
flux from our 0.5-2 keV band to the 0.3-3.5 keV EMSS band.

The second method is simple aperture photometry on the 0.5-2 keV PSPC
images of the five EMSS clusters which were targets of ROSAT
observations, i.e. in which the cluster was located at the centre of the
PSPC field. We use a large, metric aperture of 4 Mpc radius (except for
the lowest redshift cluster where we use a radius of 3 Mpc) to ensure
that almost all of the cluster flux is measured directly and corrections
for missing flux (which always require a model profile to be assumed)
remain at the less than 10\% level. The background photon list, together
with the exposure time for each photon (from the exposure map) and the
sky area associated with each photon (from its Voronoi cell), are used to
define the background level in a region outside the aperture but within
an off-axis angle of 15 arcmin. The total count rate from both source and
background photon lists within the aperture is then measured, and the
scaled background level from the background region as well as the flux
from all non-cluster sources subtracted. A small correction is made for
the cluster flux lost under non-cluster sources.

The results are given in Table 1. The WARPS method measures fluxes a mean
factor of 1.33$\pm$0.15 times higher than the EMSS method, 1.21$\pm$0.07
times higher than the Nichol et al. method, but only 1.10$\pm$0.03 times
higher than the aperture photometry method. In other words, the aperture
photometry gives fluxes that are significantly higher than those
determined in the EMSS (by a factor of 1.21) and also higher than those
determined by Nichol et al. (by a factor of 1.1).  Since, for a pure King
profile, some flux will still be outside our 4 Mpc aperture (6\% falls
outside 4 Mpc for a core radius of 250 kpc, 8\% for 350 kpc) the quoted
aperture photometry gives results consistent with the WARPS method.

We checked the aperture photometry by repeating the above procedure on
ten fields where the target was a point source (a star or AGN) of PSPC
count rate comparable to the EMSS clusters (from 0.007 count s$^{-1}$ to
0.59 count s$^{-1}$) and the exposure times were similar (8 ks to 25 ks).
The count rates measured within a 11 arcmin radius aperture,
corresponding to our 4 Mpc aperture at z$\sim0.35$, were a mean factor of
1.02$\pm$0.05 times higher than the VTP `detected' count rates,
significantly lower than the mean increase in count rate found for the
EMSS clusters (a factor of 1.11$\pm$0.03). Also, the mean VTP
`background' count rates in the apertures (i.e. including the true
background plus the cluster flux undetected by VTP) were a factor of
0.6$\pm$1.8\% lower than in the background regions in the point source
fields, compared with 12$\pm$5\% higher in the 5 EMSS cluster fields.
Thus mirror scattering in the wings of the PSF was not causing the
increased count rates in the cluster fields. In addition, a comparison
of 8 point-source fluxes measured by VTP and by Ciliegi et al.
\markcite{ci97}
(1997) using the same PSPC data gave results consistent to within 3\%.
We also used the standard ROSAT data products (the 0.5-2 keV image and
the `mex' exposure map) to measure the flux within large apertures 
for two
clusters, interactively subtracting non-cluster sources and interpolating
under them. For both MS0015.9+1609
\& MS0735.6+7421 we found a count rate within 5 Mpc which agreed with the
WARPS method to within 2\%.

The systematic difference of 20\%-30\% between the WARPS flux measurements
and the EMSS measurements explains the difference seen in the log(N)-log(S)
relations. Simple aperture photometry seems to support the WARPS
measurements.

There is a major difference in the WARPS method and that of Nichol et al.
and the EMSS. Both Nichol et al. and the EMSS assumed a fixed core radius
of 250 kpc for all clusters, whereas we estimate the core radius from the
data. The method we use to estimate the core radius is over-simplified
because it is designed for low signal-noise detections. Nevertheless, we
find a wide range of core radii within this EMSS sub-sample, from 0 kpc
to 245 kpc. For the two clusters mentioned in the introduction
(MS2137.3-2353 \& MS1512.4+3647), where HRI measurements show there are
components with small core radii (17$\pm$8 arcsec or 95 kpc and 7$\pm$1.5
arcsec or 43 kpc) we find values of 30 kpc and 80 kpc. While these
measurements may be inaccurate, or may reflect multiple components with
different spatial distributions (eg cooling flows or point sources), they
are in any case very different from 250 kpc. Thus the difference between
the WARPS fluxes on the one hand and the EMSS and Nichol et al. fluxes on
the other may result from the different treatments of the core radius.

\section{X-ray K-corrections}
Because not all of the WARPS clusters have measured redshifts, we adopt
the approach of including the K-corrections in the models.
K-corrections were calculated using for the 0.5-2 keV band using the
optically thin thermal MEKAL model spectra of Kaastra\markcite{k92} (1992) and 
Mewe et al.\markcite{m86} (1986) with
metal abundances set to 0.3 times cosmic abundance. 
The K-correction was defined here as 

\[
K_{0.5-2}={ {\int_{0.5}^{2}f_{h\nu}d(h\nu)} \over 
 {\int_{0.5(1+z)}^{2(1+z)}f_{h\nu}d(h\nu)} }
\]

where the integration limits are photon energies in keV. The results
are shown in Figure 7. For redshifts up to z=1, the K-corrections are
small ($<$20\%) for
clusters of luminosity $\sim10^{44}$ h$_{50}^{-2}$ erg s$^{-1}$, and 
thus a temperature
of $\sim$3 keV.  We include the  K-corrections
in the models by assigning a temperature to each luminosity based on
the temperature-luminosity relation of White\markcite{w96} (1996):
T(keV)=2.55x(L$_{44}$~h$_{50}^{2}$)$^{0.356}$ where L$_{44}$ is the 
X-ray luminosity
in units of 10$^{44}$ erg s$^{-1}$. This relation is valid
for luminosities of $\sim10^{43}$ h$_{50}^{-2}$ erg s$^{-1}$ to
$\sim10^{45}$ h$_{50}^{-2}$ erg s$^{-1}$ and is in reasonable agreement
with the L$_X$-T relation of Henry \& Arnaud\markcite{h91} (1991).

The dashed line in Figure 7 shows the K-corrections obtained using a
power law spectrum of energy index 0.5, as used by e.g. Henry et al. 
\markcite{h92} (1992).
Although this is a good approximation for the high temperature clusters 
more typical of the EMSS,
it systematically underestimates the flux of clusters of temperature
T$\sim$2 keV (or L$\sim$5x10$^{43}$ h$_{50}^{-2}$ erg s$^{-1}$) by
$\approx$20\% at a redshift of z=0.45, the highest redshift at which
the flux from such a cluster
would be above the WARPS flux limit.
The K-correction for clusters of even lower temperature ($\leq$1 keV) 
becomes large 
($>$1.5) at redshifts z$>$0.8, because at this temperature most of the
emission occurs at rest photon energies of $<$2 keV, and is dominated by 
iron L shell line emission at $\sim$1 keV at rest. 
In general, these large K-corrections are not needed here  because at
these  very high redshifts, the low temperature, low luminosity 
systems fall below the survey flux limit. However, the detected soft 
X-ray emission of 
clusters containing cooling flows 
may be dominated by gas at or below a temperature of 1 keV. In general,
the sensitivity of X-ray surveys (or 
at least those which use a lower energy bound $>$0.5 keV)
to high redshift cooling 
flows will be reduced by the K-correction of the cool component.

\newpage
\figcaption
{The total, corrected (unabsorbed) flux of cluster candidates (assuming
a King profile for extended sources and the instrumental PSF for
point sources) versus their raw, detected, absorbed flux. The adopted 
flux limit 
of 3.5x10$^{-14}$ \ecs (0.5-2 keV) in detected flux 
results in a flux limit of 6x10$^{-14}$ \ecs (0.5-2 keV) in
total flux.}

\figcaption
{Cluster integral Log(N)-Log(S) relation for various surveys
including WARPS (heavy filled circles). The two solid curves are
no evolution predictions for \qo=0 (upper) and \qo=0.5 (lower).
The long dashed line is
an extrapolation of the Log(N)-Log(S) relation at bright fluxes. 
The short dashed line just above the WARPS points indicates their
maximum value if all the currently unidentified `possible' candidates
are clusters.}

\figcaption
{Cluster differential Log(N)-Log(S) relation showing
the same data as in Figure 2. The solid curve is again  a 
no evolution prediction for \qo=0.5. The Mathiesen \& Evrard (1997)
models are described in the text.}

\figcaption
{High redshift (z$>$0.3) cluster differential Log(N)-Log(S)
relation. The solid curve is a no evolution prediction for \qo=0.5.
The long dashed curves are predictions based on a simple density evolution
of the XLF $\phi$(z)=$\phi(0) (1+z)^{\alpha_D}$. The short dashed
lines show the possible range of the WARPS Log(N)-Log(S).}

\figcaption
{High redshift (z$>$0.3) cluster differential Log(N)-Log(S)
relation for low luminosity ($L_X<$3x10$^{44}$ erg s$^{-1}$) clusters
only. The solid curve is a no evolution prediction for \qo=0.5,
consistent with the data.
The long dashed curves are predictions based on a simple density evolution
model as in Figure 4.}

\figcaption
{High redshift (z$>$0.3) cluster differential Log(N)-Log(S)
relation for high luminosity ($L_X>$3x10$^{44}$ erg s$^{-1}$) clusters
only. The solid curve is a no evolution prediction for \qo=0.5,
which is inconsistent with the data.
The long dashed curves are predictions based on a simple density evolution
model as in Figure 4. }

\figcaption
{X-ray K-corrections for the 0.5-2 keV band as a function of
cluster temperature as included in the model
predictions. The K-correction definition and description is given in the 
Appendix.}

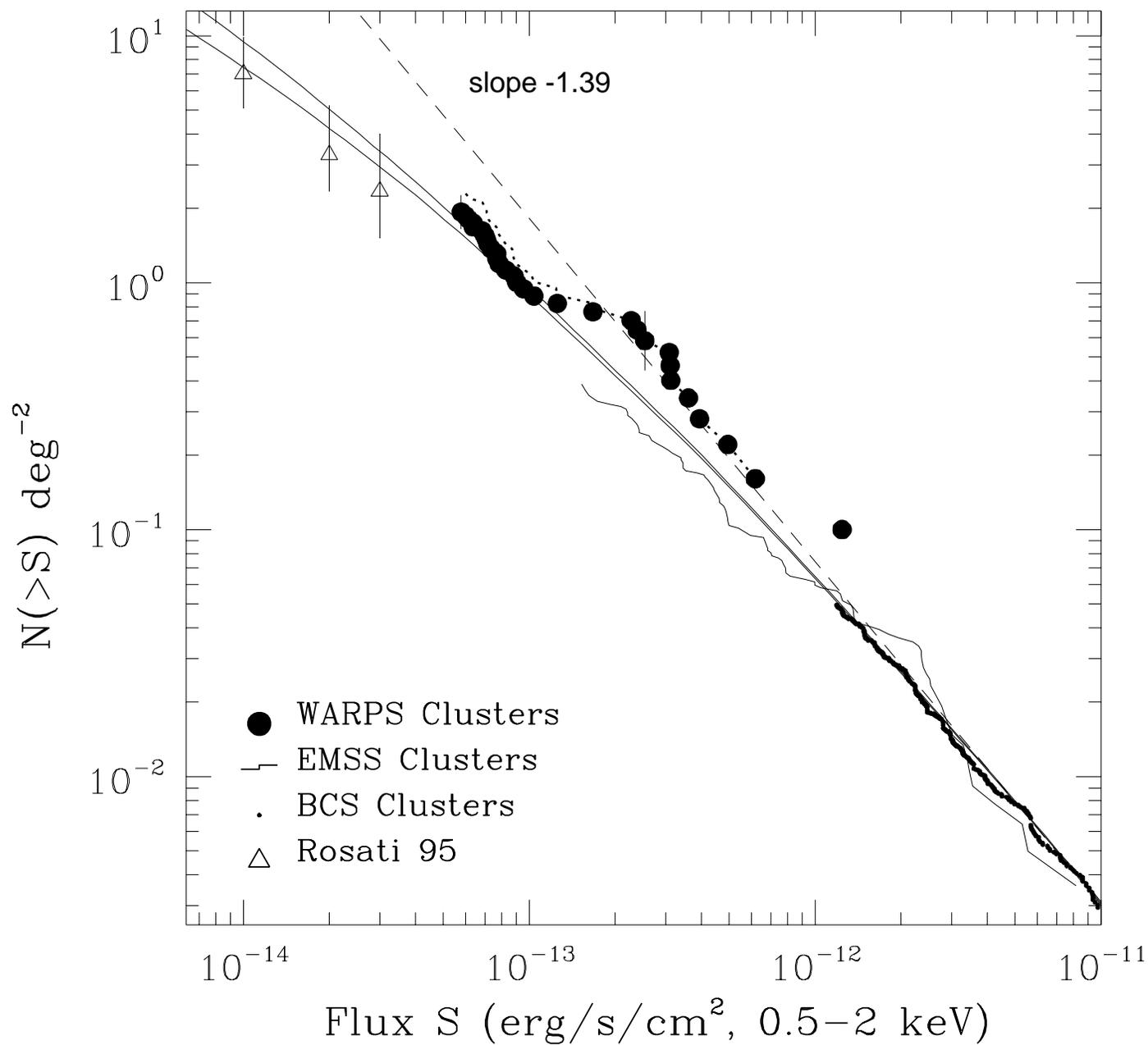
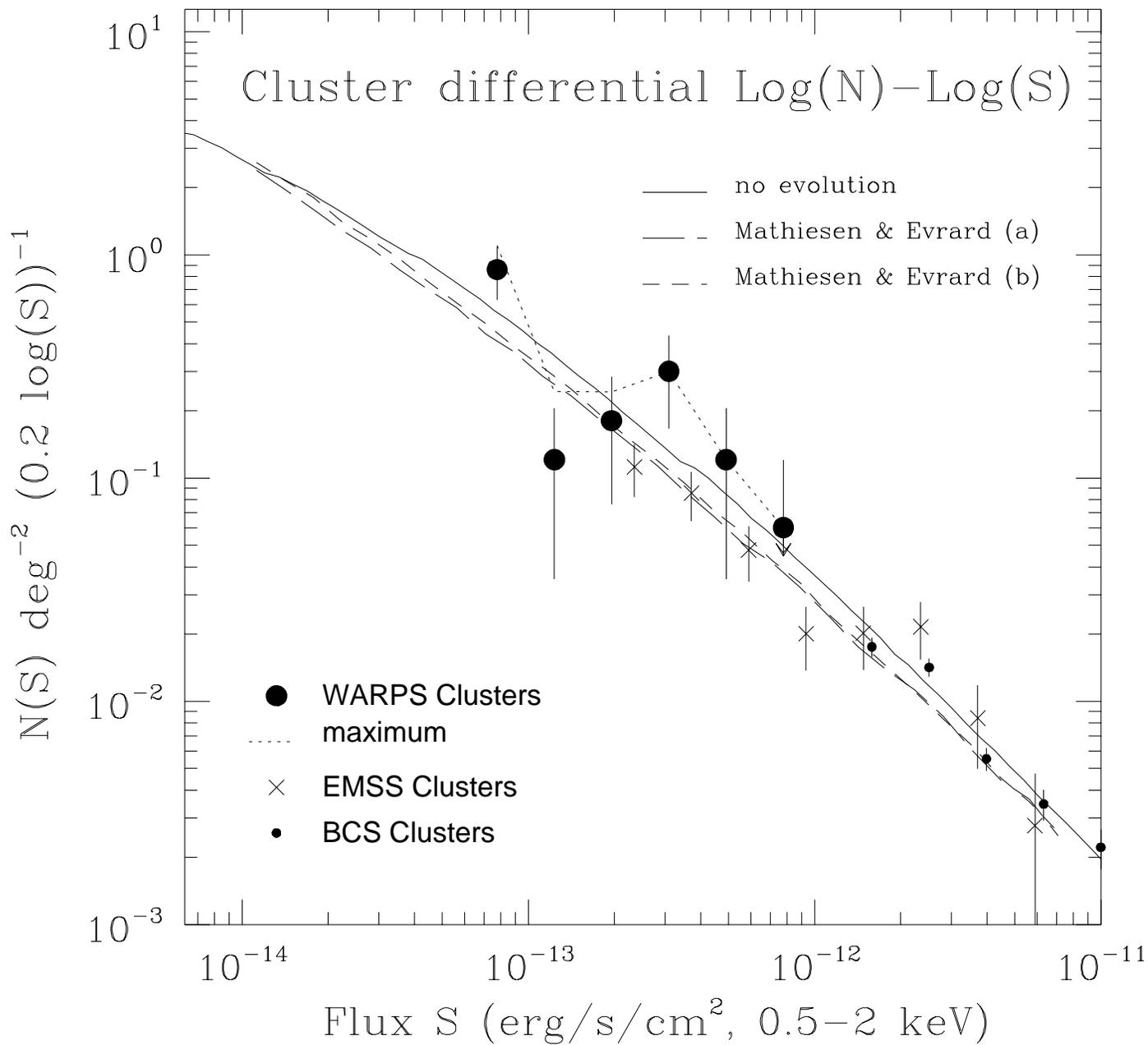
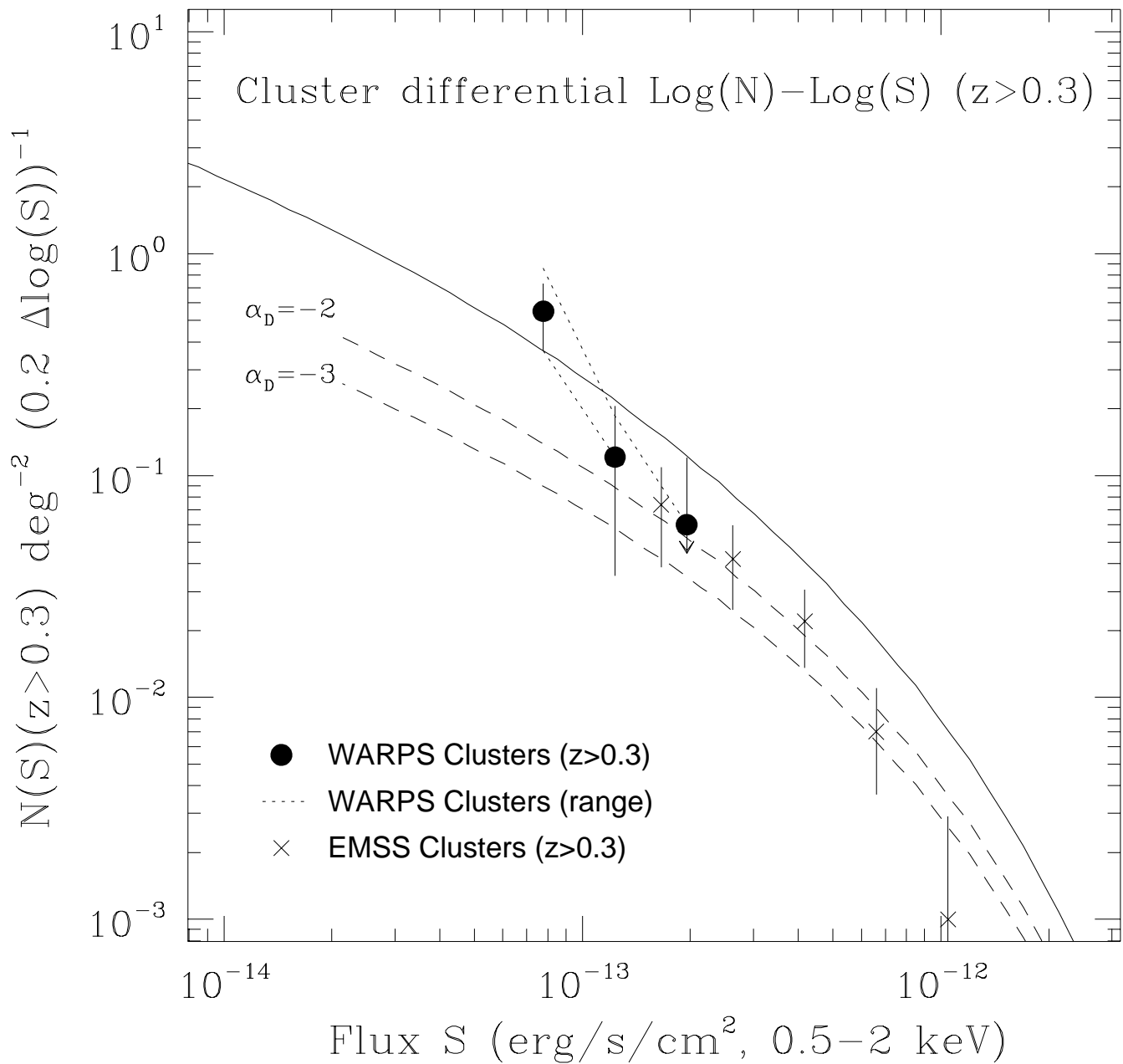
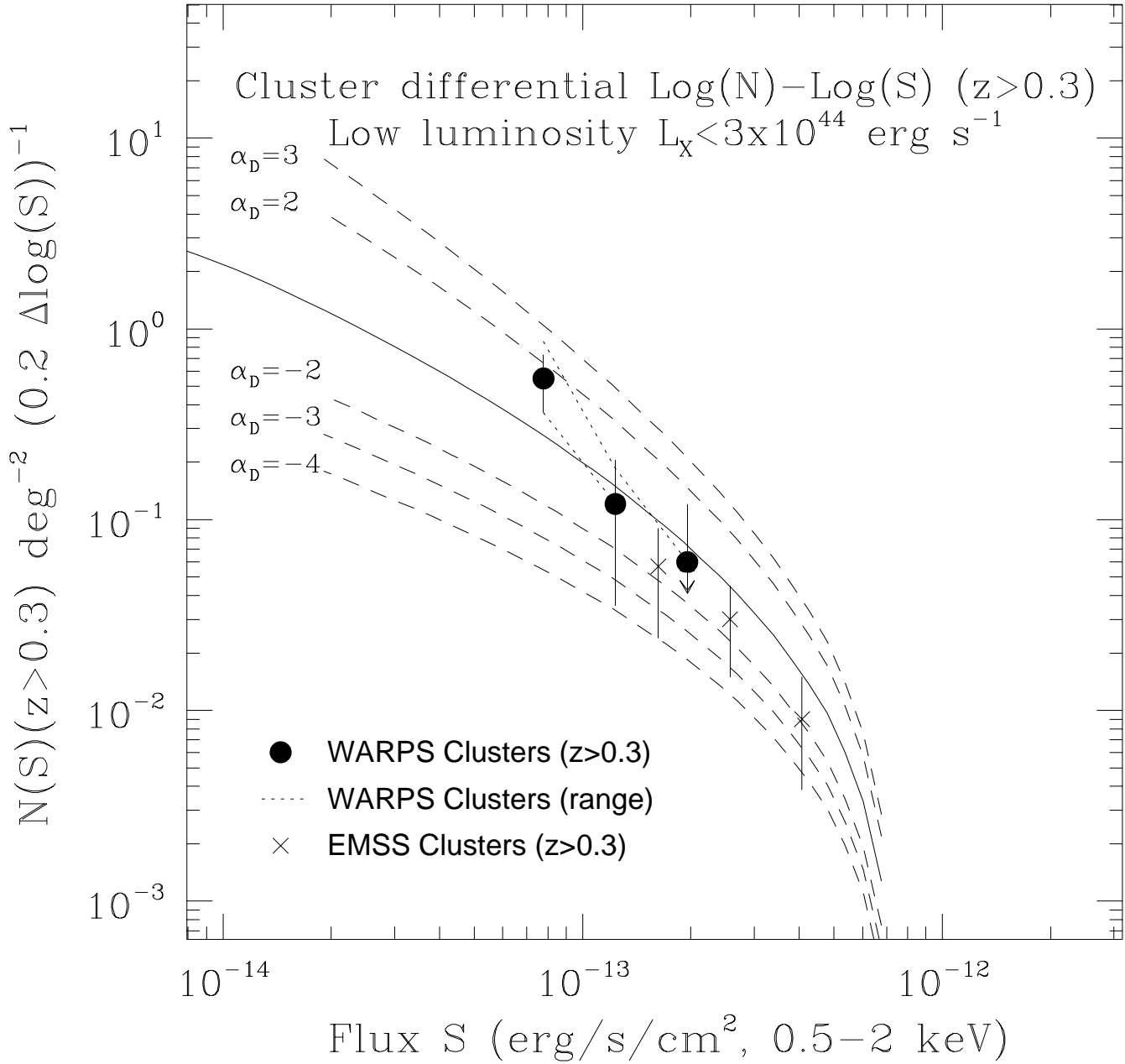
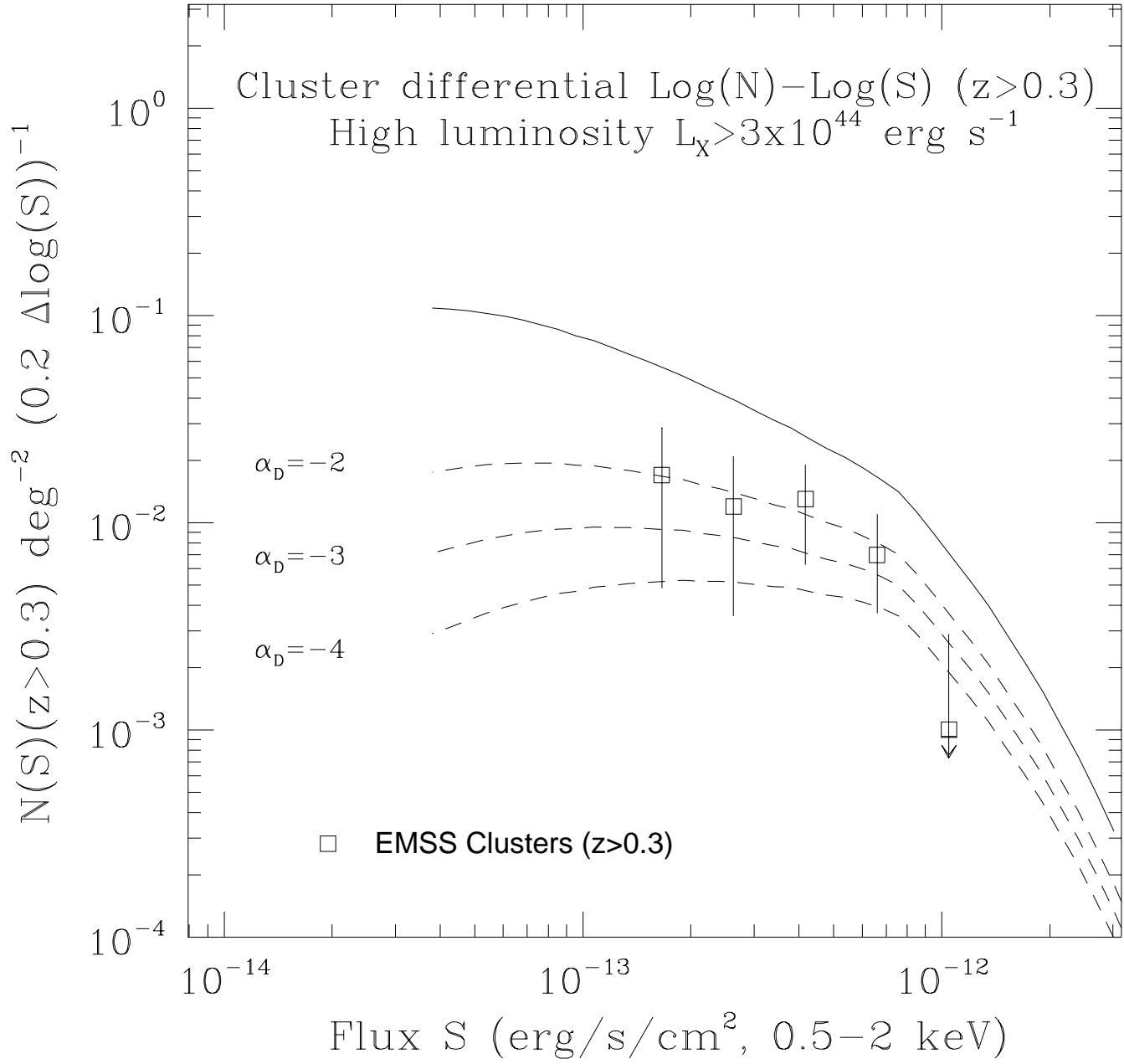
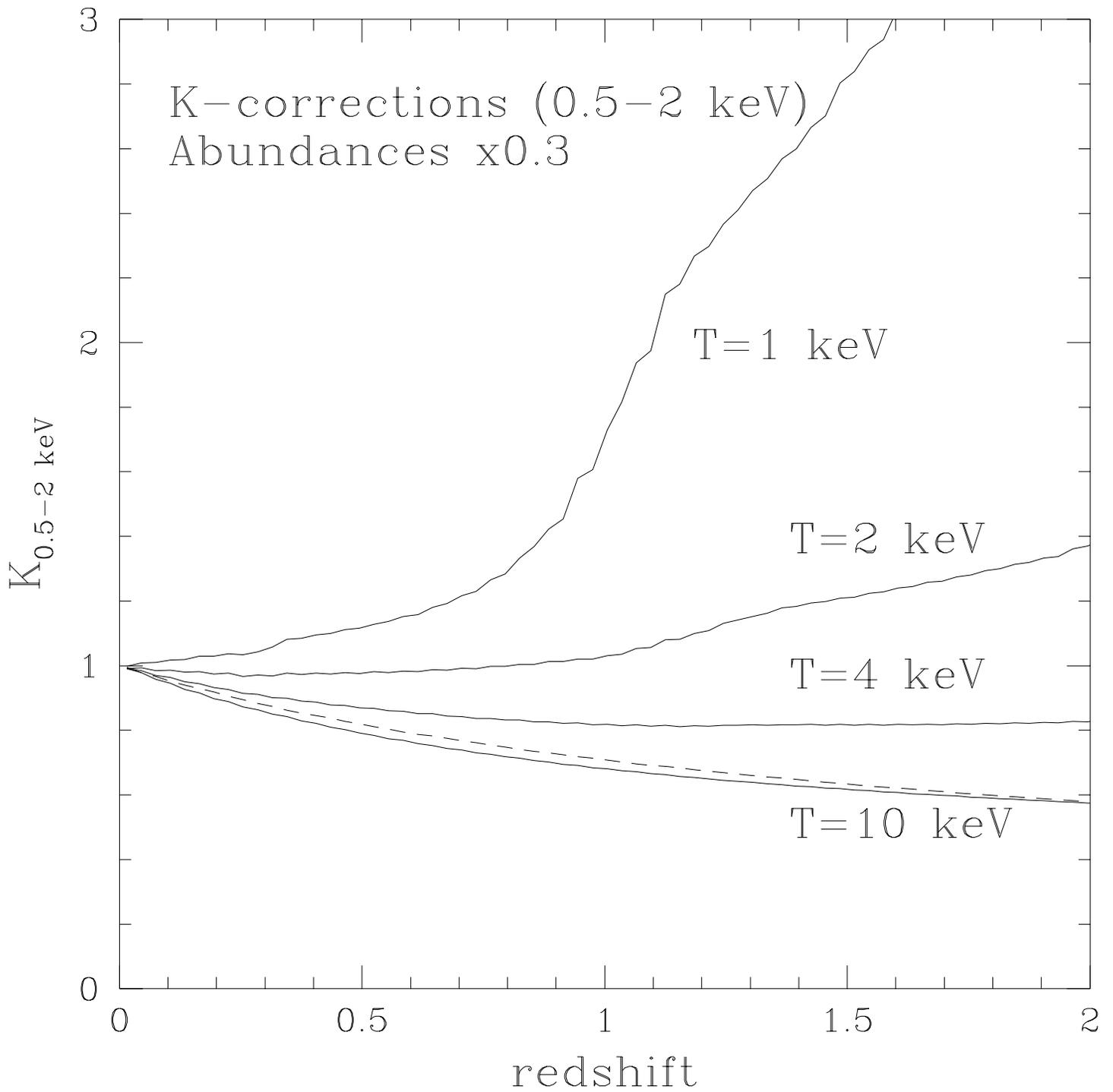
\end{document}